\documentstyle[twocolumn,prc,aps,floats,epsf]{revtex} 
\newcommand{\One}{1\kern-4.5pt1}

\newcommand{\gapprox}{\raisebox{-0.5ex}{$\ 
\stackrel{\textstyle>}{\textstyle\sim}\ $}}
\begin{document}
\preprint{SWAT/00/261}
\title{CRITICAL BEHAVIOUR IN THE DENSE PLANAR NJL MODEL}

\author{Simon Hands$^a$, Biagio Lucini$^{b,c}$ and Susan Morrison$^a$}

\address{$^a$Department of Physics, University of Wales Swansea,
Singleton Park, Swansea SA2 8PP, U.K.}

\address{
$^b$INFN Sezione di Pisa, Via Vecchia Livornese 1291,
I-56010 S. Piero a Grado (Pi), Italy.}

\address{$^c$
Department of Theoretical Physics, 1 Keble Road,
Oxford OX1 3NP, U.K.}

\twocolumn[\hsize\textwidth\columnwidth\hsize\csname@twocolumnfalse\endcsname

\maketitle

\begin{abstract}  

We present results of a Monte Carlo simulation of a 2+1 dimensional
Nambu -- Jona-Lasinio model including diquark source terms. A diquark
condensate $\langle qq\rangle$ is measured as a function of source strength
$j$. In the vacuum phase $\langle qq\rangle$ vanishes linearly with $j$ 
as expected, but simulations in a region with non-zero baryon density
suggest a power-law scaling and hence a critical system for all
$\mu>\mu_c$. There is no diquark condensation signalling superfluidity.
Comparisons are drawn with known results in two dimensional theories, 
and with the pseudogap phase in cuprate 
superconductors.
We also measure the dispersion relation $E(k)$ for fermionic excitations,
and find results consistent with 
a sharp Fermi surface. Any superfluid gap
$\Delta$ is constrained to be much less than the constituent 
quark mass scale $\Sigma_0$.
\end{abstract}


]

In the absence of a reliable non-perturbative method for calculating
the properties of QCD at high densities, models of strongly-interacting matter
continue to rely on results derived from effective theories such as the 
Nambu -- Jona-Lasinio (\hbox{NJL}) 
model \cite{NJL}. Recently such approaches have
suggested the intriguing possibility of superconducting behaviour in quark
matter at high density as a result of BCS condensation between diquark pairs 
at the Fermi surface \cite{ARW,ARW2}. Even in simplified models,
however, systematic methods of calculation are hard to find, and up to now
mostly self-consistent 
methods have been applied \cite{BR,PR,BRS}. Scenarios such as 
color-flavor locking \cite{ARW2}, however, imply that in the high
density phase the global U(1)$_B$ symmetry in the QCD Lagrangian
corresponding to conservation of baryon number is spontaneously
broken, leading to superfluid behaviour and a massless scalar excitation via
Goldstone's theorem. A description of the fluctuations of these modes and the 
resulting long-range interactions may need a more systematic non-perturbative
approach; in this Letter we will use numerical results from lattice field
theory to argue that the situation is more complex than hitherto thought.

The Lagrangian for the NJL model, incorporating a baryon chemical
potential $\mu$, is 
\begin{equation}
{\cal L}_{NJL}=\bar\psi({\partial\!\!\!/}+m+\mu\gamma_0)\psi
-{g^2\over2}\left[(\bar\psi\psi)^2-(\bar\psi\gamma_5\vec\tau\psi)^2\right],
\end{equation}
where $\psi,\bar\psi$ carry a global isopsin index acted on by 
Pauli matrices $\tau_i$.
The model in 2+1 dimensions has been studied with $\mu\not=0$ using
staggered lattice fermions $\chi,\bar\chi$ \cite{HM}: 
the reader is referred to this paper
for further details of the formulation and symmetries of the lattice model. 
Apart from the obvious numerical advantages of working in a reduced
dimensionality, the model in this case
also has an interacting continuum limit for 
a critical value of the coupling $1/g^2_c\simeq1$
\cite{GNstuff}, making controlled predictions
possible in principle. In the simulations of this Letter, 
as in \cite{HM}, $1/g^2=0.5$,
rather far from the continuum limit. For $\mu=0$, this results in a ground state
with broken chiral symmetry and a dynamically generated fermion mass
$\Sigma_0\simeq0.71$, together with a triplet of Goldstone pions.
The ``constituent quark mass'' $\Sigma_0$ defines the physical scale
in cutoff units.
As $\mu$ is raised from zero, the system remains essentially unaltered 
until $\mu=\mu_c\simeq0.65$, whereupon the constituent quark mass $\Sigma$
falls to zero as chiral symmetry is restored 
in a strong first-order transition, and baryon density
$n=\langle\bar\psi\gamma_0\psi\rangle$ jumps discontinuously from zero to 
a non-zero value; by  $\mu=0.8$, the density 
$n\simeq0.25$ quarks of each isopsin component
per lattice site.

The tendency for diquark pairing at high density
was investigated via measurements
of the diquark propagator $G(t)=\sum_{\bf x}\langle qq(0,{\bf0})
\bar q\bar q(t,{\bf x})\rangle$ \cite{HM}. In particular it was found that
for $\mu>\mu_c$ and the choice of isoscalar diquark operator
$qq=\chi^{tr}(x)\tau_2\chi(x)$ (the ``spectral scalar'' channel of \cite{HM}),
$G(t)$ plateaus at a non-zero value as $t\to\infty$, implying 
$\langle qq\rangle\langle\bar q\bar q\rangle\not=0$ by the cluster property.
Studies on varying spatial volumes, however, reveal that the
plateau height is not an extensive quantity as naively expected, and that 
diquark condensation signalling a superfluid state
is not unambiguously indicated.

The situation can be clarified by the introduction of source terms for 
diquark and anti-diquark pairs:
\begin{equation}
{\cal L}[j,\bar\jmath]={\cal L}_{NJL}+j\chi^{tr}(x)\tau_2\chi(x)
             +\bar\jmath\bar\chi(x)\tau_2\bar\chi^{tr}(x).
\end{equation}
The functional integral may now be written using the {\sl Gor'kov}
representation \cite{HM2} as
\begin{equation}
Z[j,\bar\jmath]=
\langle{\rm Pf}({\cal A}[j,\bar\jmath])\rangle,
\label{eq:meas}
\end{equation}
where $\langle\ldots\rangle$ denotes averaging with respect to bosonic auxiliary
fields $\sigma$ and $\vec\pi$ introduced to make ${\cal L}_{NJL}$ bilinear
in $\chi$ and $\bar\chi$, and the antisymmetric matrix ${\cal A}$
is 
\begin{equation}
{\cal A}=\left(\matrix{\bar\jmath\tau_2&{1\over2}M\cr
         -{1\over2}M^{tr}&j\tau_2\cr}\right),
\end{equation}
where $M=M(\mu;\sigma,\vec\pi)$ 
is the conventional kinetic operator of ${\cal L}_{NJL}$ \cite{HM}.
For convenience we now define diquark operators $qq_\pm$ via
\begin{equation}
qq_\pm(x)={1\over2}\left[\chi^{tr}(x)\tau_2\chi(x)
\pm\bar\chi(x)\tau_2\bar\chi^{tr}(x)\right],
\end{equation}
with corresponding sources 
$j_\pm=j\pm\bar\jmath$. 
The diquark condensate is readily calculated to be
\begin{equation}
\langle qq_+\rangle={1\over V}{{\partial\ln Z}\over{\partial j_+}}
={1\over{4V}}\langle{\rm tr}\,\tau_2{\cal A}^{-1}\rangle.
\end{equation}
Under U(1)$_B$ transformations
$\chi\mapsto e^{i\beta}\chi$, $\bar\chi\mapsto\bar\chi e^{-i\beta}$, the
operators $qq_\pm$ rotate into each other; condensation of $\langle
qq_+\rangle\not=0$ implies a massless Goldstone mode in the $qq_-$ channel
as $j_+\to0$,
illustrated by a Ward identity for the susceptibility $\chi_-$:
\begin{equation}
\chi_-\equiv\sum_x\langle qq_-(0)qq_-(x)\rangle\biggr\vert_{j_-=0}
={{\langle qq_+\rangle}\over{j_+}}.
\label{eq:Ward}
\end{equation}

Numerical simulation using the functional measure (\ref{eq:meas}) requires
us to address the issue of calculating the Pfaffian \cite{CEGL}. We begin by 
observing that ${\rm Pf}{\cal A}=\pm{\rm det}^{1\over2}{\cal A}$, 
the sign depending on the details of the ordering of the Grassmann
integration. Now, using the property of a square block matrix 
\begin{equation}
{\rm det}\left(\matrix{X&Y\cr W&Z\cr}\right)=
{\rm det}X{\rm det}(Z-WX^{-1}Y),
\end{equation}
together with the property $\tau_2M\tau_2=M^*$,
we find ${\rm det}{\cal A}={\rm det}(j\bar\jmath+M^\dagger M/4)$, and hence
is real and positive if $j\bar\jmath$ is chosen real and positive.
It follows that ${\rm Pf}{\cal A}$ is real.
To avoid the sign problem, however, we choose to simulate
with measure ${\rm det}^{1\over2}({\cal A^\dagger A})={\rm Pf}^2{\cal A}$,
corresponding to a doubling in the number of fermion species. This 
is usual practice in simulations of four-fermi models; ``crosstalk''
between the two species, which might conceivably lead to problems for 
$\mu\not=0$ due to light unphysical baryonic states in the spectrum, is
subleading in this case and causes no problems \cite{Barb}. It is 
straightforward to implement the simulation using a hybrid molecular 
dynamics `R' algorithm \cite{R} --- the resulting dynamics
are those of 2 flavors of staggered lattice fermion, corresponding to
$N_f=4$ flavors of four-component physical fermion.

\begin{figure}[htb]
\begin{center}
\vspace*{-.5cm}
\epsfxsize=3.2in
\epsfbox{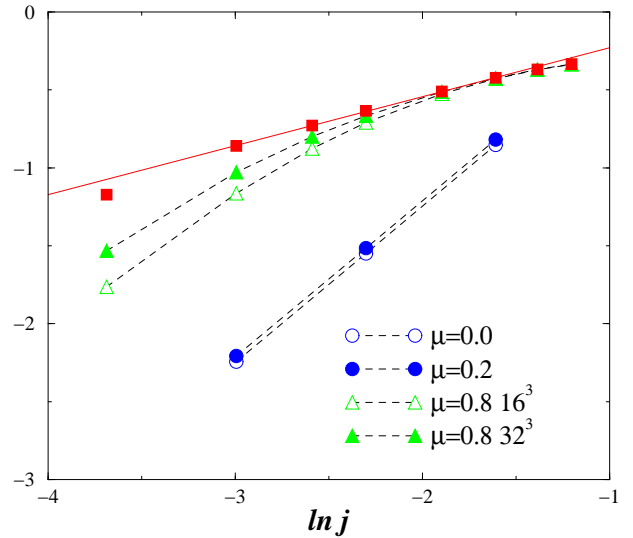}
\end{center}
\vspace{-.3cm}
\caption{
\label{fig:lnln}
$\ln\langle qq_+\rangle$ vs. $\ln j$}
\vspace{-.35cm}
\end{figure}
\begin{figure}[htb]
\begin{center}
\vspace*{-.5cm}
\epsfxsize=3.2in
\epsfbox{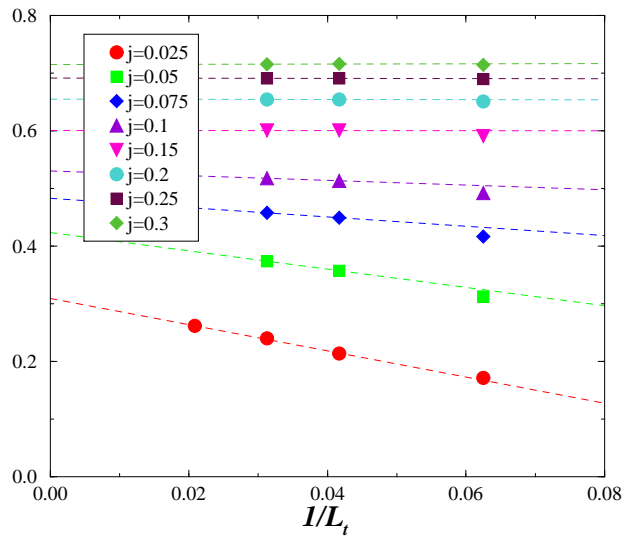}
\end{center}
\vspace{-.3cm}
\caption{
\label{fig:extrapolate}
Extrapolation of $\langle qq_+\rangle$  at $\mu=0.8$ 
to the infinite volume limit.}
\vspace{-.35cm}
\end{figure}

We have performed simulations on lattice sizes $16^3$, $16^2\times24$, $24^3$,
$32^3$ and $48^3$, with chemical potentials $\mu=0,0.2$ corresponding
to the low density chirally-broken phase, and $\mu=0.8$ in the high density
chirally-restored phase \cite{HM}. Diquark source values ranging from 
$j=0.3$ down to as low as $j=0.025$ were used; throughout we took 
$\bar\jmath=j$
and kept a bare 
Dirac mass $m=0.01$ to assist identification of chirally broken and restored
phases. 
We plot $\ln\langle qq_+\rangle$ vs. $\ln j$ 
in Fig.~\ref{fig:lnln}. 
At low
density the results for $\mu=0$ and $\mu=0.2$ are very similar and support
a linear relation $\langle qq_+\rangle\propto j$, suggesting that 
U(1)$_B$ is unbroken in the zero-source limit. 
For $\mu=0.8$, when chiral
symmetry is restored, the situation is different: whilst there is still
no evidence for a non-zero condensate as $j\to0$,
the log-log plot now displays a marked 
curvature, together with evidence for significant finite volume
effects.
Empirically we find that on a $L_s^2\times L_t$ system the
dominant correction scales as $1/L_t$ -- this enables an
extrapolation of the data with $L_t\geq24$ to the thermodynamic limit, shown in 
Fig.~\ref{fig:extrapolate}.

The extrapolated data, denoted by filled squares in Fig.~\ref{fig:lnln},
suggest the best description
in the high-density phase is a power law: 
\begin{equation}
\langle qq_+\rangle\propto j^\alpha,
\label{eq:powerlaw}
\end{equation}
with $\alpha\simeq0.31$. There is thus
no superfluid condensate in this region
of parameter space, and no consequent Goldstone pole. 
The Ward identity (\ref{eq:Ward}), however, implies
that the transverse susceptibility $\chi_-$
diverges as $j^{\alpha-1}$; thus
long-wavelength fluctuations remain important, and there are
strongly-interacting massless states present. 
This is the first main result 
of this Letter.

Now, the value $\mu=0.8$ was originally chosen merely as characteristic of the 
high-density phase; moreover data from partially quenched simulations
\cite{HM2} suggest that the power-law scaling
(\ref{eq:powerlaw}) may be generic for $\mu>\mu_c$. This is analogous to the
behaviour of the $2d$ X-Y model \cite{IDZJ}, 
which in accordance with well-known theorems
\cite{Mermin}
does not have spontaneous magnetisation, but instead exhibits a line
of critical points for all $T<T_c$ \cite{KT}. In this low-temperature phase
the modulus of the spin variables is constant, but the phase fluctuates; the
divergence of $\chi_-$ signals correlation of the phase
over macroscopic distance
scales {\sl without\/} long-range phase coherence. This critical behaviour
without an order parameter can be characterised by two exponents $\delta$
and $\eta$, defined in our notation by $\langle qq_+\rangle\propto
j^{1\over\delta}$, and $\lim_{j_+\to0}\langle
qq_-(0)qq_-(x)\rangle\propto1/x^{d-2+\eta}$. For the X-Y model \cite{IDZJ,KT}
$\delta\geq15$
and $\eta\leq{1\over4}$ (the inequalites being saturated up to logarithmic
corrections at $T=T_c$), and satisfy the hyperscaling relation
\begin{equation}
\delta={{d+2-\eta}\over{d-2+\eta}}.
\label{eq:hyper}
\end{equation}
For the dense planar NJL model, by contrast, our results suggest
$\delta\simeq3$.

The phenomenon of long-range phase coherence being wiped out by soft
transverse fluctuations is particularly interesting for a {\sl composite\/}
order parameter field, suggesting as it does that some 
dynamical pairing mechanism is in play even in the absence of symmetry breaking.
It has been suggested that such a state may describe a {\sl pseudogap}
phase, characterised by a suppression of spectral weight 
in the vicinity of the Fermi
surface, and
observed as a precursor to the superconducting state for
underdoped cuprate superconductors \cite{Randeria}.
It is intriguing in the current context 
that cuprate superconductivity is a planar phenomenon
\footnote{An important distinction is that in cuprates
the pairing operator is $d$-wave, whereas for the NJL model it is $s$-wave.}.

A field theoretic example of possible relevance is studied in 
\cite{BigEd}, where results from an exactly soluble $2d$
fermionic model are generalised to 
the Gross-Neveu model with U(1) chiral symmetry \cite{GN}.
Once again due to the low dimensionality no long-range ordering 
via $\langle\bar\psi\psi\rangle\not=0$ is possible; however for sufficiently
large number of flavors $N_f$ the model has a phase analogous to the 
low temperature phase of the X-Y model, with $\eta\sim O(1/N_f)$.
The physical fermion is a superposition of positive and negative chirality
states and has zero net chirality;
despite the absence of chiral symmetry breaking it
propagates as $C(x)\propto x^{-({1\over2}+\eta)}e^{-\Sigma x}$, implying
massive propagation as $x\to\infty$, but also 
a non-trivial spectral function,
signalled by the departure from canonical
scaling of the pre-exponential factor.

\begin{figure}[htb]
\begin{center}
\vspace*{-.5cm}
\epsfxsize=3.2in
\epsfbox{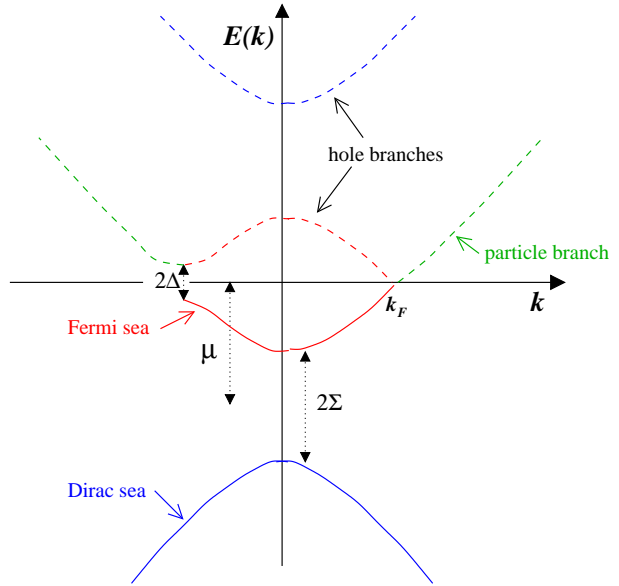}
\end{center}
\vspace{-.3cm}
\caption{
\label{fig:disperse}
Dispersion relation $E(k)$, for gapped (left)
and ungapped (right) cases.}
\vspace{-.35cm}
\end{figure}
These considerations motivate a study of the spectrum 
of spin-${1\over2}$ excitations in the NJL model
by analysing the fermion propagator
\hbox{$C({\bf k},t)=\sum_{\bf x}{\rm Re}\langle
\chi(0,{\bf0})\bar\chi(t,{\bf x})e^{i{\bf k.x}}\rangle$}, where for staggered
fermions the sum only includes {\bf x} which are an even number 
of lattice spacings from the origin in each direction.
Note that for $j\not=0$, 
the most general definition of $C(t)$ contains both ``normal''
$\langle\chi\bar\chi\rangle$ and ``anomalous'' $\langle\chi\chi\rangle$
components. In this Letter we analyse only the isosinglet normal
component, with {\bf k} oriented along a lattice axis. 
Fig.~\ref{fig:disperse} shows 
possible spectra $E(k)$ for $\mu>\mu_c$, $n>0$ \cite{PR}. Solid lines denote 
states which are occupied in the ground state; following common usage we label 
these ``Dirac'' and ``Fermi'' seas (note there is no quantum number which
distinguishes between these branches). 
Dashed lines denote empty states, and 
hence describe the spectrum of allowed excitations. 
Note that for $k<k_F$ the lowest excitations vacate states in the
Fermi sea, and hence are `hole-like': in contrast for $k>k_F$, 
excitations
add quarks to the system, and are `particle-like'.
On the right hand side
the curve is continuous as it crosses the axis; there is
a sharp Fermi surface with well-defined Fermi momentum $k=k_F$, where 
excitations cost zero energy. On the left hand side we have sketched the 
effect of a discontinuity at the Fermi surface arising
from a BCS instability;
the lowest excitation energy now corresponds to the gap $\Delta$.

\begin{figure}[htb]
\begin{center}
\vspace*{-.5cm}
\epsfxsize=3.2in
\epsfbox{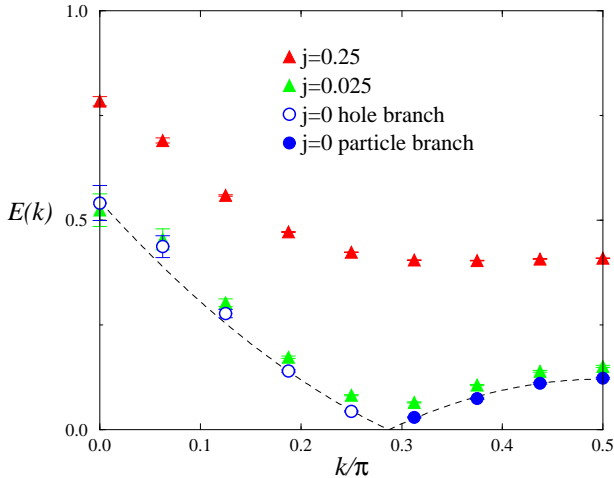}
\end{center}
\vspace{-.3cm}
\caption{
\label{fig:e(k)}
$E(k)$ at $\mu=0.8$.}
\vspace{-.35cm}
\end{figure}
We have measured $C({\bf k},t)$ on $32^3$ lattices at $\mu=0.8 $. Empirically we
find that for $t$ even it is consistent with zero, and that for $t$ odd it is
well described by a 3-parameter fit
$C({\bf k},t)=Ae^{-Et}+Be^{-E(L_t-t)}$.
For small values of $k$ we find $A\gg B$, ie. the signal predominantly
forwards-moving, and vice-versa for $k$ as approaches
its maximum value $\pi/2$. As $j$ is decreased the switch between these two
behaviours becomes increasingly pronounced. On comparison with the analytic form
for free fermions, we label these cases as `hole' and `particle'
respectively. We also find that $E(k,j)$ decreases monotonically and 
roughly linearly with $j$, and a linear extrapolation to $j=0$ can be
made based on
$j\in[0.1,0.025]$. The results, shown in Fig.~\ref{fig:e(k)}, are in 
good qualitative agreement with the expectations of Fig.~\ref{fig:disperse}. 
Conservatively, we can bound any possible gap from above, and find
$\Delta<0.05\ll\Sigma_0\sim0.7$, in contrast with expectations based on 
the self-consistent gap equation approach of \cite{BR}. More boldly, we plot 
a curve of the form $E(k)=\vert P+Q({\pi\over2}-k)^2\vert$, which
suggests it is plausible the dispersion relation is continuous
near $k=k_F$, and 
hence $\Delta=0$. The fit yields a $k_F\simeq0.90$ slightly larger
than the Fermi energy $\mu$, which is characteristic of a system with 
attractive inter-particle forces.
We conclude that there is a sharp Fermi surface in this region of parameter
space: this is the second main result of the Letter.

In the future, it will be helpful to explore other values of $\mu>\mu_c$, 
to see whether $\delta=\delta(\mu)$. Independent estimates of $\eta$ from 
simulations at $j=0$ would also be valuable; one could then use
hyperscaling (\ref{eq:hyper}) to probe the effective dimensionality
of the system. It has been suggested that absence of diquark 
condensation in the planar NJL model is an artifact due to
the low dimensionality of the Fermi surface in 2+1 dimensions \cite{HM}; 
it should 
be mentioned, however, that partially quenched 
data taken on small 3+1 dimensional systems 
also appear to scale according to (\ref{eq:powerlaw}) \cite{HM2}.
Finally, by analogy with cuprate superconductivity, it may be the case that 
this critical phase we have found is a precursor to a superfluid phase setting
in at higher densities. Simulations with the current parameters become
plagued by lattice artifacts for $\mu\gapprox1.0$ \cite{Barb}. To observe
superfluid behaviour simulations closer to 
the continuum limit may be necessary. 

\vspace{-12pt}
\acknowledgements

This work is supported
by the TMR-network ``Finite temperature phase transitions in particle physics''
EU-contract ERBFMRX-CT97-0122; in addition
SJH thanks the Institute for Nuclear Theory
at the University of Washington for its hospitality and the 
Department of Energy and the Leverhulme Trust
for partial support.
We were helped by discussions with Mark Alford, Nick Dorey, Helen Fretwell, 
Tim Hollowood,
Krishna Rajagopal, Dirk Rischke, Misha Stephanov, Costas 
Strouthos and Jac Verbaarschot.

\end{document}